\documentstyle[aps,pra]{revtex}

\begin{document}
\title{Quantum computation with two-level trapped cold ions
beyond Lamb-Dicke limit}
\author{L. F. Wei, S. Y. Liu and X. L. Lei}
\address{Department of Physics, Shanghai Jiao Tong University,
1954 Huashan Road, Shanghai 200030, China}

\maketitle
\begin{center}
(Received: 31 July 2001)
\end{center}

\begin{abstract}
We propose a simple scheme for implementing quantum logic gates
with a string of two-level trapped cold ions outside the
Lamb-Dicke limit. Two internal states of each ion are used as one
computational qubit (CQ) and the collective vibration of ions acts as
the information bus, i.e., bus qubit (BQ). Using the quantum dynamics for the
laser-ion interaction as described by a generalized Jaynes-Cummings model,
we show that quantum entanglement between any one CQ and the BQ can be
coherently manipulated by applying classical laser beams. As a result,
universal quantum gates, i.e. the one-qubit rotation and two-qubit controlled
gates, can be implemented exactly. The required experimental parameters
for the implementation, including the Lamb-Dicke (LD) parameter and
the durations of the applied laser pulses, are derived.
Neither the LD approximation for the laser-ion interaction nor the
auxiliary atomic level is needed in the present scheme.

\vspace{0.3cm}

PACS number(s): 89.70.+c, 32.80. Qk, 03.67.Lx.
\end{abstract}

\vspace{0.5cm}

\section{Introduction}

Since Shor's algorithm for efficiently factoring large numbers
was proposed\cite{Shor}, many authors have addressed the problem of implementing quantum computation\cite{EJ}.
It has been shown that arbitrary rotations in the Hilbert space of
an individual computational qubit (CQ), i.e., one-qubit gate,
and a controlled rotation, such as a controlled-NOT ($C^N$) or
controlled-Z ($C^Z$), between two different CQs,
i.e., a two-qubit controlled gate, are universal quantum gates.
In other words, any unitary transformation on arbitrarily many CQs can be carried out by
repeatedly performing these universal quantum gates \cite{Bar1}.
Since one-qubit gates are generally easy to realize,
implementing the two-qubit controlled gate operation
is a central problem for constructing a quantum computer.
Several kinds of physical systems
have been proposed to implement quantum logic operations.
Some of the most attractive schemes are based on nuclear magnetic resonance (NMR) \cite
{GC}, coupled quantum dots \cite{LD}, cavity QED \cite{Tru}, and 
trapped cold ions \cite{Ste}, introduced first by Cirac and Zoller\cite{CZ1}.
This system is based on the laser-ion
interaction and possesses long qubit coherence times \cite{KS}.
Information is stored in the spin states of an array of trapped cold ions
and manipulated by using laser pulses. Two special conditions are required
in the original Cirac-Zoller scheme \cite{CZ1}: a) the ion
must have three levels, and b) the trap must operate in the Lamb-Dicke (LD) limit, i.e.,
the coupling between the internal spin and external vibrational
degrees of freedom of the ion must be sufficiently weak.
Laser pulses with different
polarizations are used to coherently manipulate the
quantum information of the system and five-step pulses are used
to realize an exact $C^N$ logic operation between two different cold ions.
It is thus desirable to look for simpler schemes to construct ion-trap
quantum computer.

In the last few years, a number of modifications and extensions to Cirac-Zoller's idea
have been proposed \cite{Tur2,Mor2,Lei,CC}.
In particular, carrying out the universal quantum logic gates
with a single trapped cooled ion had been paid much attention, because 
imprisoning a single ion in a trap and cooling it to the vibrational ground
state can be easily achieved with the present experimental technology. In
1995, Monroe $et$ $al$ \cite{Mor1} demonstrated a simplified scheme  for
realizing the "reduced" two-qubit $C^N$ logic operation between the internal
and external degrees of freedom of the trapped cold ion by applying three
laser pulses. Furthermore, by using a single resonant pulse, a two-qubit
controlled operation with a single trapped two-level ion outside the LD
limit was constructed \cite{Mor2}. However, the third
auxiliary internal atomic state is still required in Ref.\,\cite{Mor1} and
the operation reported in Ref.\,\cite{Mor2} is not the exact $C^N$ gate,
although it is equivalent to the "reduced" $C^N$ logic gate \cite{Mor1}
apart from phase factors. Subsequent operations must be
carried out in order to eliminate the additional phase factors.
Up to now, the problem of construction a quantum computing network by connecting
different ions in different traps, i.e., the scalability of the single
trapped ion, has not been solved \cite{CZ2}.
On the other hand, much attention has been paid to the problem of building a quantum network
with cold ions confined in a single trap.
Recent experiments showed that the collective motion of two $^{9}$Be ions
confined in a trap can be cooled to the ground state \cite{King} and that
quantum entanglement of up to four ions can be been achieved\cite{Sac},
thus supporting the idea of realizing the quantum computation in a single
trap. Theoretically, Jonathan {\em et al} \cite{Jon}
proposed an alternative scheme to realize the true two-qubit ion-ion gate
by using the AC Stark shift induced by laser resonance with the ionic
transition frequency. This scheme allows an increase
in gate speed by at least an order of magnitude with respect to
that of related single ion-trap experiments \cite{Mor1,Roo}.
In addition, directly modifying the Cirac-Zoller's proposal \cite{CZ1},
Childs and Chuang \cite{CC} provided a general and accessible technique
for performing universal logic operation between ions with only
two internal levels in a common trap.
However, these schemes base on the LD approximation,
which requires that the spatial dimension of the ground state of the
collective motion of these ions is much smaller than
the effective wavelength of the laser wave.
In fact, the quantum motion of the trapped ions is not limited to the LD
regime \cite{Ste}. It has been shown that utilizing the laser-ion interaction
beyond the LD limit is helpful for reducing the noise in the trap and
improving the cooling rate \cite{Stev}. 

In this paper we propose an alternative scheme for realizing quantum
computation with a set of two-level cold ions in a single trap driven by
a series of laser pulses without using the LD approximation. The CQ is
encoded by two internal states of the ion and the collective vibration
of the trapped ions acts as the information bus, i.e., the bus qubit (BQ).
We show how to realize universal quantum gates, including a simple
rotation on any individual CQ and the exact two-qubit $C^Z$ or $C^N$ logic 
operation, by using suitable laser pulses.
The paper is arranged as follows: In Sec.\,II the conditional quantum
dynamics for the laser-ion interaction are derived and the method for
coherently manipulating the entanglement between the CQ and BQ is proposed.
Then, we present a simpler scheme to realize rotations
on an individual CQ and the exact two-qubit controlled logic operations
between the CQ and the BQ. We derive values of the required parameters, including the LD parameter
and the durations of the applied laser pulses.
Sec.\,III is devoted to the construction of logic gates between
different CQs by making use of the elementary operations presented in
Sec.\,II. We also discuss how to set up the experimental parameters for these
realizations. Finally, we give some conclusions in Sec.\,IV.

\section{Quantum dynamics for the laser-ion interaction beyond LD limit and 
exact quantum gates with a single trapped cold ion}

\vspace{0.2cm} The use of a BQ makes the physical construction of a
quantum information processor much simpler. Most current ion trap
proposals use this idea. The BQ carries the quantum information in
the computer. Instead of seeking a means to carry out the quantum
logic operation between a pair of CQs directly, it is
sufficient to implement quantum operation between an arbitrary CQ and the BQ.
In an ion-trap quantum computer a CQ is encoded by two internal spin states
of the ion and the BQ is encoded by two Fock states of the external collective
vibration of the trapped ions (center-of-mass vibrational degree of freedom).
It has been assumed that a single laser beam can be directed
at will to any chosen ion \cite{CZ1}.
Thus, a single trapped cold ion driven by a laser beam can be used
to describe a single CQ interacting with the BQ. In this section we show 
how to implement quantum gates with a single trapped cold ion driven by
a travelling-wave laser field by considering the conditional quantum dynamics
for the laser-ion interaction without taking LD approximation.

For simplicity, we assume that a single ion is stored in a coaxial
resonator RF-ion trap\cite {Jef}, which provides pseudopotential oscillation
frequencies satisfying the condition $\omega _{x}<<\omega _{y,z}$ along
the principal axes of the trap. Only the quantized vibrational motion
along $x$ direction is considered for the cooled ion\cite{Mor1}.
Following Blockley {\em et al }\cite{Blo}, the
interaction between a single two-level trapped cold ion and a classical
single-mode travelling light field of frequency $\omega _{L}$ can be
described by the following Hamiltonian:
\begin{eqnarray}
\hat{H}(t)=\hbar \omega (\hat{a}^{\dagger }\hat{a}+\frac{1}{2})+\frac{1}{2}%
\hbar \omega _{0}\hat{\sigma}_{z}+\frac{\hbar \Omega }{2}\{\hat{\sigma}%
_{+}\exp [i\eta (\hat{a}+\hat{a}^{\dagger })-i(\omega _{L}t+\phi )]+H.c.\}.
\end{eqnarray}
The first two terms correspond to the ion's external and internal degrees of freedom, respectively, and $\omega $ is the trap frequency.
The final term gives the interaction between the ion and the light field with phase $\phi $.
Pauli operators $\hat{\sigma}_{z}$ and $\hat{\sigma}_{\pm }$ describe the
internal degrees of freedom of the ion. $\hat{a}^{\dagger }$ and $\hat{a}$
are the creation and annihilation operators of the trap vibrational quanta.
$\omega _{0}$ is the atomic transition frequency, and $\Omega$ is the
Rabi frequency. $\eta (<1)$ is the LD parameter. We consider the case in which the applied 
laser is resonant with the
$k$th red-shifted vibrational sideband, i.e., the frequency of laser field
is chosen as $\omega_{L}=\omega _{0}-k\omega $, where $k$ is a positive integer. Under the usual
rotating-wave approximation, the Hamiltonian of the system reads
\begin{eqnarray}
\hat{H}=\frac{\hbar \Omega }{2}\exp (-\frac{\eta ^{2}}{2}-i\phi)\{\hat{\sigma}%
_{+}(i\eta )^{k}[\sum_{n=0}^{\infty }\frac{(i\eta )^{2n}\hat{a}^{\dagger n}%
\hat{a}^{n}}{n!(n+k)!}]\hat{a}^{k}+H.c.\},
\end{eqnarray}
in the interaction picture defined by the unitary operator $\hat{U}%
_{0}(t)=\exp [-i\omega t(\hat{a}^{\dagger }\hat{a}+1/2)]\exp (-it\delta
\hat{\sigma }_{z}/2)$, where $\delta =\omega _{0}-\omega _{L}$ is the
detuning of laser field with the ion. The above Hamiltonian is similar to
that of nonlinear coupled multiquantum Jaynes-Cummings model \cite{Vog},
which is exactly solvable. Therefore, it is easy to check the dynamical
evolution of any two-qubit initial state by using evolution operator
$\hat{U}(t)=\exp(-\frac{i}{\hbar }\hat{H}t)$.
Indeed, with the help of relation  \cite{Stea}
$$
\langle m-k|\langle e|\hat{H}|m\rangle|g\rangle=\left\{
\begin{array}{l}
0 ,\qquad m<k; \\
\\
\\
\hbar i^k e^{-i\phi}\Omega _{m-k,m},\qquad m\geq k,
\end{array}
\right.
$$
with 
$$
\Omega _{m-k,m}=\frac{\Omega \eta ^{k}e^{-\eta ^{2}/2}}{2}\sqrt{\frac{(m)!}{(m-k)!}%
}\sum_{n=0}^{m-k}\frac{(i\eta)^{2n}}{(k+n)!}C^n_{m-k},
$$
the time evolution of the initial
states $|m\rangle |e\rangle $ and $|m\rangle|g\rangle$ can be expressed as
\begin{eqnarray}
|m\rangle |e\rangle \longrightarrow \cos \Omega _{m,m+k}t|m\rangle |e\rangle
-(-i)^{k-1}e^{i\phi }\sin \Omega _{m,m+k} t|m+k\rangle |g\rangle ,
\end{eqnarray}
and
\begin{eqnarray}
|m\rangle |g\rangle \longrightarrow \left\{
\begin{array}{l}
|m\rangle |g\rangle ,\qquad m<k;\vspace{0.2cm} \\
\\
\cos \Omega _{m-k,m}t|m\rangle |g\rangle +i^{k-1}e^{-i\phi }\sin \Omega
_{m-k,m}t|m-k\rangle |e\rangle ,\qquad m\geq k,
\end{array}
\right.
\end{eqnarray}
respectively. The above treatment can also be modified directly to another
laser excitation case, i.e., the $k$th blue-sideband. In the present work only the red-sideband excitation is considered.
It is seen from equations (3) and (4) that entangled states are produced by
the time evolution of the state $|m\rangle |e\rangle $ and the state
$|m\rangle |g\rangle $ with $m\geq k.$ From this conditional quantum dynamics we can 
define two kinds of elementary quantum operations: the one-qubit rotations  
\begin{eqnarray}
\hat{r}_c(m,\phi,t)=\{\cos \Omega _{m,m}t|g\rangle \langle g|-ie^{-i\phi }\sin \Omega _{m,m}t
|e\rangle \langle g|-ie^{i\phi }\sin \Omega _{m,m}t|g\rangle \langle
e|+\cos \Omega _{m,m}t|e\rangle \langle e|\}\otimes |m\rangle\langle m|,
\end{eqnarray}
generated by applying a resonant 
pulse ($\omega_L=\omega_0$) to the chosen ion, and the two-qubit joint operation on the CQ and BQ 
\begin{eqnarray}
\hat{R}_{cb}(m,\phi,t)=\left\{
\begin{array}{ll}
|m\rangle|g\rangle\langle m|\langle g|+
(\cos\Omega _{m,m+k}t|m\rangle|e\rangle-
(-i)^{k-1}e^{i\phi}\sin\Omega _{m,m+k}t|m+k\rangle|g\rangle)\langle m|\langle e|, \hspace{0.2cm}m<k;\\
\\
\\
(\cos\Omega_{m-k,m}t|m\rangle|g\rangle+
i^{k-1}e^{-i\phi}\sin\Omega _{m-k,m}t|m-k\rangle|e\rangle)\langle m|\langle g|\\
\\
\hspace{2.7cm}+(\cos\Omega _{m,m+k}t|m\rangle|e\rangle-
(-i)^{k-1}e^{i\phi}\sin\Omega _{m,m+k}t|m+k\rangle|g\rangle)\langle m|\langle e|, \hspace{0.2cm}m\geq k,
\end{array}
\right.
\end{eqnarray}  
performed by using an off-resonant pulse. Here $\phi$ and $t$ are the
initial phase and duration of the applied pulse.
By making use of these basic operations we now show how to realize  
exact quantum logic operations on a single trapped ion,   
once the LD parameter and the laser-ion interaction time are set
up properly. We only need to consider the cases $k=0$ and $k=1$.

\subsection{Simple rotations of a single CQ}

In general, one-qubit rotations are easy to implemented in a
physical system for quantum computation. In fact, it has shown  that,
under the LD approximation, the simple rotation of a CQ can be realized
directly by applying a resonant pulse ($\omega_L=\omega_0$) on the
specifically chosen ion \cite{CZ1,CC,Sas}.
This operation does not depend on the state of the BQ.
The reason is that, under the LD approximation, the applied resonant pulse
does not result in a coupling between the internal and external degrees
of freedom of the ion. Beyond the LD limit, however, such a coupling exists
and thus the rotating angle of the one-qubit operation (5) depends on
both the state of the BQ and the duration of the applied resonant pulse.
In the space spanned by the basis states $\Gamma =\{|0\rangle,
|1\rangle\}\otimes\{|g\rangle, |e\rangle\}$,
the transformation $\hat{r}_c(m,\phi, t)$ takes the following matrix
form:
\begin{eqnarray}
\hat{r}_{c}(\phi, t)=\left(
\begin{array}{cccc}
cos\Omega _{0,0}t & -ie^{i\phi }sin\Omega _{0,0}t & 0 & 0 \\
-ie^{-i\phi }sin\Omega _{0,0}t & cos\Omega _{0,0}t & 0 & 0 \\
0 & 0 & cos\Omega _{1,1}t & -ie^{i\phi }sin\Omega _{1,1}t \\
0 & 0 & -ie^{-i\phi }sin\Omega _{1,1}t & cos\Omega _{1,1}t
\end{array}
\right) .
\end{eqnarray}
Here $|0\rangle$ and $|1\rangle$ are the ground and first excited
states of the external vibration of the trapped ion. They are usually used
to encode the BQ. It is easily seen that the Walsh-Hadamard gate 
can also be implemented by this one-qubit rotation. Indeed, 
the resonant pulse with special initial phase $\phi$
and the duration $t=(\pi/4+2k\pi)/\Omega _{m,m}$
yields the following operation on the CQ,
\begin{eqnarray}
\hat{r}_c(m, \phi,t)\longrightarrow\{
\begin{array}{l}
\frac{1}{\sqrt{2}}|m\rangle\langle m|\otimes
\left(
\begin{array}{cc}
1&1\\
-1&1
\end{array}
\right )\hspace{3mm} 
{\mbox for} \hspace{2mm}\phi=\pi/2\pm 2k\pi, k=0,1,2...,\\
\\
\frac{1}{\sqrt{2}}|m\rangle\langle m|\otimes
\left(
\begin{array}{cc}
1&-1\\
1&1
\end{array}
\right )\hspace{3mm}
{\mbox for}\hspace{2mm} \phi=3\pi/2\pm 2k\pi,
\end{array}
\end{eqnarray}
which generates a 
uniform superposition of two encoded basis states of the CQ from one of them.
The state of the BQ is unchanged 
during this operation. It is easily seen that, if the condition
$$
\cos \Omega _{0,0}t=1, \hspace{3mm} \sin\Omega _{1,1}t=1,
$$
is satisfied, the operation (7) reduces to the controlled operation
constructed in Ref.\,\cite{Mor2,LG}, which is equivalent to the $C^N$
logic operation between the BQ and a CQ under local transformation.

\subsection{Logic operation between the CQ and BQ:\,\,$C^Z$ gate}

It is easily seen that an off-resonant pulse is required for generating
entanglement between the CQ and BQ from an unentangled two-qubit
initial state. We assume in what follows that the cold ion is
addressed by the first red-sideband laser pulse, i.e. the frequency of
the applied laser field is $\omega_L=\omega_0-\omega$. The BQ is encoded
by the states $|0\rangle$ and $|1\rangle$. As a consequence, the dynamical evolution 
equation (6) can be rewritten as
$$
\hat{R}_{cb}(0,\phi,t)=|0\rangle|g\rangle\langle 0|\langle g|
+(\cos\Omega_{0,1}t|0\rangle|e\rangle-e^{i\phi}\sin\Omega_{0,1}t|1\rangle|g\rangle)
\langle 0|\langle e|,
$$
for the BQ's initial state $|0\rangle$, and 
$$
\hat{R}_{cb}(1,\phi,t)=(\cos\Omega_{0,1}t|1\rangle|g\rangle+
e^{-i\phi}\sin\Omega_{0,1}|0\rangle|e\rangle)\langle 1|\langle g|
+(\cos\Omega_{1,2}t|1\rangle|e\rangle-e^{i\phi}\sin\Omega_{1,2}t|2\rangle|g\rangle)
\langle 1|\langle e|,
$$
for the BQ's initial state $|1\rangle$. Obviously, if the applied  
off-resonant pulse satisfies the condition
\begin{eqnarray}
\cos \Omega _{0,1}t=1,\hspace{1cm}\cos \Omega _{1,2}t=-1,
\end{eqnarray}
a universal two-qubit logic gate, the controlled-Z ($\hat{C}^Z_{cb}$)
logic operation between the CQ and BQ in $\Gamma$ space, i.e.,
\begin{eqnarray}
\hat{C}^Z_{cb}=|0\rangle |g\rangle \langle 0\langle g|+|0\rangle |e\rangle
\langle 0|\langle e|+|1\rangle |g\rangle \langle 1|\langle g|-|1\rangle
|e\rangle \langle 1|\langle e|=\left(
\begin{array}{cccc}
1 & 0 & 0 & 0 \\
0 & 1 & 0 & 0 \\
0 & 0 & 1 & 0 \\
0 & 0 & 0 & -1
\end{array}
\right) ,
\end{eqnarray}
can be realized directly. If the BQ is in the state $|0\rangle$,
$\hat{C}^Z_{cb}$ has no effect, whereas if the BQ is in the state
$|1\rangle$, $\hat{C}^Z_{cb}$ rotates the state of the CQ by the Pauli
$\sigma_z$ operator. Similarly, one can prove that if the duration
of the applied pulse satisfies the condition
$$
\cos \Omega _{0,1}t^{\prime }=1,\hspace{1cm}\sin \Omega _{1,2}t^{\prime }=1,
$$
the $\hat{C}^Z_{cb}$ logic operation (10) can also be implemented
by sequentially applying two red-sideband pulses with equal durations
$t^{\prime }=2p^{\prime }\pi /\Omega _{0,1},p^{\prime }=1,2,3,...$, i.e.,
\begin{eqnarray}
\left\{
 \begin{array}{l}
|0\rangle |g\rangle \longrightarrow |0\rangle |g\rangle
\longrightarrow |0\rangle |g\rangle ,\hspace{2mm}
|0\rangle |e\rangle \longrightarrow |0\rangle |e\rangle
\longrightarrow |0\rangle |e\rangle , \\
\\
|1\rangle |g\rangle \longrightarrow |1\rangle |g\rangle
\longrightarrow |1\rangle |g\rangle ,\hspace{2mm}
|1\rangle |e\rangle \longrightarrow -|2\rangle |g\rangle
\longrightarrow -|1\rangle |e\rangle .
\end{array}
\right.
\end{eqnarray}
However, in the following,
only the direct way to implement the $\hat{C}^Z_{cb}$ gate (10) is
considered. It is worth noting that there is no requirement on the
initial phase of the applied pulse for realizing the
operation $\hat{C}^Z_{cb}$.

\subsection{Logic operation between CQ and BQ: \,\,$C^N$ gate}

The "reduced"
$\hat{C}^N_{cb}$ logic gate\cite{Mor1} can be constructed exactly from
the one-qubit operation $\hat{r}_{c}(m,\phi,t)$
and the two-qubit gate $\hat{C}^Z_{cb}$. Indeed, one can 
easily see that the operation $\hat{C}^Z_{cb}$, surrounded by two one-qubit 
operations $\hat{r}_c(\phi_1,t_1)$ and $\hat{r}_c(\phi_3,t_3)$, 
yields the following new operation between the BQ and a CQ:
\begin{eqnarray}
\hat{r}_{c}(\phi_3,t_{3})\hat{C}^Z_{cb}\hat{r}_{c}(\phi_1,t_{1})=\left(
\begin{array}{cccc}
A_{0000} & A_{0001} & 0 & 0 \\
A_{0100} & A_{0101} & 0 & 0 \\
0 & 0 & A_{1000} & A_{1011} \\
0 & 0 & A_{1110} & A_{1111}
\end{array}
\right) ,
\end{eqnarray}
with
$$
%\begin{eqnarray}
\left\{
\begin{array}{l}
A_{0000}=\cos \Omega _{0,0}t_{3}\cos \Omega _{0,0}t_{1}-e^{i(\phi _{3}-\phi
_{1})}\sin \Omega _{0,0}t_{3}\sin \Omega _{0,0}t_{1}, \\
\\
A_{0001}=-ie^{i\phi _{1}}\cos \Omega _{0,0}t_{3}\sin \rho
_{0}t_{1}-ie^{i\phi _{3}}\sin \Omega _{0,0}t_{3}\cos \Omega _{0,0}t_{1}, \\
\\
A_{0100}=-ie^{-i\phi _{3}}\sin \Omega _{0,0}t_{3}\cos \rho
_{0}t_{1}-ie^{-i\phi _{1}}\cos \Omega _{0,0}t_{3}\sin \Omega _{0,0}t_{1}, \\
\\
A_{0101}=\cos \Omega _{0,0}t_{3}\cos \Omega _{0,0}t_{1}-e^{-i(\phi
_{3}-\phi _{1})}\sin \Omega _{0,0}t_{3}\sin \Omega _{0,0}t_{1}; \\
\\
A_{1010}=\cos \Omega _{1,1}t_{3}\cos \Omega _{1,1}t_{1}+e^{i(\phi
_{3}-\phi _{1})}\sin \Omega _{1,1}t_{3}\sin \Omega _{1,1}t_{1}, \\
\\
A_{1011}=-ie^{i\phi _{1}}\cos \Omega _{1,1}t_{3}\sin \rho
_{1}t_{1}+ie^{i\phi _{3}}\sin \Omega _{1,1}t_{3}\cos \Omega _{1,1}t_{1}, \\
\\
A_{1110}=ie^{-i\phi _{1}}\cos \Omega _{1,1}t_{3}\sin \rho
_{1}t_{1}-ie^{-i\phi _{3}}\sin \Omega _{1,1}t_{3}\cos \Omega _{1,1}t_{1}, \\
\\
A_{1111}=-\cos \Omega _{1,1}t_{3}\cos \Omega _{1,1}t_{1}-e^{-i(\phi
_{3}-\phi _{1})}\sin \Omega _{1,1}t_{3}\sin \Omega _{1,1}t_{1},
\end{array}
\right.
%\end{eqnarray}
$$
where $t_{1}, t_{3}$ and $\phi_1,\phi_3$ are the durations
and initial phases of the first and third applied resonant laser pulses, respectively.
They should be set up properly for realizing the specific quantum operation
between the BQ and CQ. It is easily seen that,
if the two initial phases $\phi_1,\phi_3$ satisfy the relation
\begin{equation}
\phi_3-\phi_1=\pm 2k\pi,\hspace{0.2cm}k=0,1,2,...,
\end{equation}
the matrix elements on the right side of Eq.\,(12) become
$$
%\begin{eqnarray}
\left\{
\begin{array}{l}
A_{0000}=A_{0101}=\cos \Omega _{0,0}(t_{3}+t_1),\hspace{2mm}
A_{0001}=-ie^{i\phi _{1}}\sin \Omega _{0,0}(t_{3}+t_1), \hspace{0.2cm}
A_{0100}=-ie^{-i\phi _{1}}\sin \Omega _{0,0}(t_{3}+t_1), \\
\\
A_{0101}=-A_{1111}=\cos \Omega _{0,0}(t_{3}-t_1), \hspace{2mm}
A_{1011}=ie^{i\phi _{1}}\sin \Omega _{1,1}(t_3-t_{1}),\hspace{0.2cm}
A_{1110}=-ie^{-i\phi _{1}}\sin \Omega _{1,1}(t_3-t_{1}).
\end{array}
\right.
%\end{eqnarray}
$$
Furthermore, if the matching conditions
\begin{eqnarray}
\cos \Omega _{0,0}(t_{3}+t_{1})=1,\hspace{0.3cm}
\sin \Omega _{1,1}(t_{3}-t_{1})=\left\{
\begin{array}{ll}
1 & \mbox{for $\phi_1=3\pi/2\pm 2k'\pi, k'=0,1,2,...;$} \\
\\
-1 & \mbox{for $\phi_1=\pi/2\pm 2k'\pi, k=0,1,2,...,$}
\end{array}
\right .
\end{eqnarray}
are satisfied, we have 
$$
A_{0000}=A_{0101}=A_{1011}=A_{1110}=1,\hspace{0.3cm}
A_{0001}=A_{0100}=A_{1000}=A_{1111}=0.
$$
This means that, under the conditions (13) and (14), the operation (12) is
nothing but the exact reduced $C^N$ logic gate \cite{Mor1}, i.e.,
\begin{eqnarray}
\hat{r}_{c}(\phi_3,t_{3})\hat{C}^Z_{cb}\hat{r}_{c}(\phi_1,t_{1})\longrightarrow \left(
\begin{array}{cccc}
1 & 0 & 0 & 0 \\
0 & 1 & 0 & 0 \\
0 & 0 & 0 & 1 \\
0 & 0 & 1 & 0
\end{array}
\right) =\hat{C}^N_{cb}.
\end{eqnarray}
In this logic operation, if the BQ (control qubit) is in the
state $|0\rangle$, the operation has no effect, whereas if the BQ is
in the state $|1\rangle$, a NOT operation is applied to the CQ.
The BQ remains in its initial state after the operation.

We now show how to set up the experimental parameters to realize
the $\hat{C}^N_{cb}$ gate (15). These parameters include the LD parameter
and the durations of the applied laser pulses. First, the
requisite LD parameter and the duration of the off-resonant  
red-sideband pulse are determined by equation (9), which implies
\begin{eqnarray}
\frac{\Omega _{1,2}}{\Omega _{0,1}}=\frac{2-\eta ^{2}}{\sqrt{2}}=\frac{q-0.5%
}{p},\,\,p,q=1,2,3,...,\hspace{0.3cm}t_{2}=\frac{2p\pi }{\Omega _{0,1}}=\frac{4p\pi e^{\eta ^{2}/2}}{\Omega \eta}.
\end{eqnarray}
Second, using the LD parameter determined from the above equation,
the durations of two resonant pulses surrounding the first red-sideband
pulse are further determined by solving equation (15) for
different initial phases, e.g.,
\begin{eqnarray}
\left\{
\begin{array}{l}
t_{1}=\pi (\frac{p^{\prime }+1}{\Omega _{0,0}}+\frac{q^{\prime }+0.25}{\Omega _{1,1}}%
)=T_1,\hspace{0.2cm}t_{3}=\pi (\frac{p^{\prime }+1}{\Omega _{0,0}}-\frac{q^{\prime
}+0.25}{\Omega _{1,1}})=T_3,\hspace{0.2cm}{\rm for}\hspace{0.2cm}\phi _{1}=\phi
_{3}=\pi /2; \vspace{0.3cm}\\
\\
t_{1}=T_3,\hspace{0.2cm}t_{3}=T_1,\hspace{0.2cm}{\rm for}\hspace{0.2cm}\phi _{1}=\phi
_{3}=3\pi /2;\hspace{0.2cm}p^\prime,q^\prime=1,2,3....
\end{array}
\right.
\end{eqnarray}

So far we have shown that, once the coupling parameter
$\eta $ is set up appropriately, the "reduced" two-qubit $C^N$ logic operation
between the CQ and BQ can be implemented exactly by sequentially applying
three pulses (a red-sideband pulse surrounded by two resonant ones)
with controllable durations.
As in Refs.\,\cite{Mor2,LG},
neither the weak coupling limit (i.e. LD approxiamtion) nor the auxiliary
atomic level is required in the present scheme.
Therefore, pulses having different polarizations are not needed
and the coupling parameter between the CQ and BQ may be large
(e.g. $0.9064$, $0.9692$, $etc.$). We also note that this logic operation
does not depend on the initial phase of the applied off-resonant pulse,
while the initial phases of two resonant pulses should be set up accurately
to satisfy the conditions (13-14).
The main points of our approach can be
summarized as follows:\vspace{0.2cm}

a). use a resonant pulse with initial phase $\pi/2$(or $3\pi/2$) and
duration $T_1$($T_3$) to rotate the chosen CQ,\vspace{0.1cm}

b). use an off-resonant red-sideband pulse to complete the $C^Z$ logic
operation between the chosen CQ and BQ, \vspace{0.1cm}

c). use another resonant pulse with initial phase $\pi/2$(or $3\pi/2$) and 
duration $T_3$ ($T_1$) on the chosen CQ to complete the gate
operation $C^N$ between the chosen CQ and BQ.\vspace{0.2cm}

Some values of these experimental parameters are given in Table I.
The durations of the applied pulses are given by the quantities
$\Omega t_{j}/\pi, j=1,2,3$ in the table. It is seen from the table
that the switching speed of the $C^N$ gate depends on the Rabi frequency
$\Omega $ and the LD parameter $\eta$. It can be estimated numerically
once the experimental parameters are defined. For example,
in the case of of a recent experiment \cite{Mor1} a single $^{9}$Be$^{+}$ ion
confined in a coaxial-resonator radio frequency (RF)-ion trap and
cooled to its quantum ground state by Raman cooling, the target qubit
consists of two $^{2}S_{1/2}$ hyperfine ground states of $^{9}$Be$^{+}$:
$%
^{2}S_{1/2}|F=2,m_{F}=2\rangle$ and $^{2}S_{1/2}|F=2,m_{F}=1\rangle$,
separated by 
$\omega
_{0}/{2\pi }=1.25$\,GHz. If the Rabi frequencies are chosen as $\Omega =2\pi \times 140$\,kHz for resonant excitations and $\eta
\Omega =2\pi \times 30$\,kHz for off-resonant excitations, one can easily
show that the shortest duration of the applied pulses for the realization
of the exact $C^N$ gate is about $10^{-4}$\,second. 

\begin{center}
TABLE I. Some experimental parameters for realizing the exact "reduced" $C^N$ gate
by three-step sequential pulses with $\phi_1=\phi_3=\pi/2$.
\end{center}
\begin{center}
\begin{tabular}{|c|c|c|c|c|c|c|c|}
\hline
$p$ & $q$ & $\eta$ & $\Omega t_2/\pi$ & $p^{\prime}$ & $q^\prime$ & $\Omega t_1/\pi $ & $\Omega t_3/\pi $\\
\hline
$2$ &$2$&$       $ & $13.2024  $ & $5   $ & $1  $ & $29.1785   $ & $2.8108   $\\
\cline{5-8}
$10$ &$8$&$0.9692$ & $33.0061  $ & $8   $ & $1  $ & $38.7753   $ & $12.4076  $\\
\cline{5-8}
$\cdots$ &$\cdots$&$  $ & $\cdots  $ & $10  $ & $1  $ & $45.1732   $ & $18.8055  $\\
\cline{1-8}

$2$ &$3$&$     $ & $18.6448  $ & $1   $ & $1  $ & $2.9777    $ & $1.5148    $\\
\cline{1-2}\cline{4-4}\cline{5-8}
$6$ &$8$&$0.4819$ & $55.9343  $ & $2   $ & $2  $ & $8.1496    $ & $0.8354    $\\
\cline{1-2}\cline{4-4}\cline{5-8}
$\cdots$&$\cdots$& $ $ &$\cdots$ & $3   $ & $1  $ & $7.4702    $ & $6.0073    $\\
\cline{1-8}

$3$ &$3$&$     $ & $19.9648  $ & $2   $ & $1  $ & $10.2554    $ & $1.8081    $\\
\cline{1-2}\cline{4-4}\cline{5-8}
$9$ &$8$&$0.9064$ & $59.8943  $ & $3   $ & $1  $ & $13.2713  $ & $4.8239    $\\
\cline{1-2}\cline{4-4}\cline{5-8}
$\cdots$&$\cdots$& $ $ &$\cdots$ & $8   $ & $2  $ & $45.2453    $ & $3.0088  $\\
\cline{1-8}

$4$ &$6$&$0.2355$ & $69.8532 $ & $1   $ & $1  $ & $2.6005 $ & $1.5119   $\\
\cline{5-8}
$$ &$$&$$ & $  $ &               $2   $ & $1  $ & $4.6567 $ & $3.5682  $\\
\cline{5-8}
$\cdots$&$\cdots$& $ $ &$\cdots$ & $2   $ & $2  $ & $6.8337 $ & $1.3913 $\\
\cline{1-8}

$14$ &$20$&$0.1738$ & $16.3571 $ & $1   $ & $1  $ & $2.5538    $ & $1.5071 $\\
\cline{5-8}
$$ &$$&$$ & $  $ & $2   $ & $1  $ & $4.5843  $ & $3.5374    $\\
\cline{5-8}
$\cdots$&$\cdots$& $ $ &$\cdots$ & $2   $ & $2  $ & $6.6779$ & $1.4438 $\\
\cline{1-8}

$3$ &$4$&$0.5919$ & $24.1611 $ & $1   $ & $1  $ & $3.2991    $ & $1.4661 $\\
\cline{5-8}
$$ &$$&$$ & $  $ & $2   $ & $1  $ & $5.6817  $ & $3.8487    $\\
\cline{5-8}
$\cdots$&$\cdots$& $ $ &$\cdots$ & $2   $ & $2  $ & $9.3477$ & $0.1827 $\\
\cline{1-8}
$\cdots$&$\cdots$& $ $ &$\cdots$ & $\cdots   $ & $\cdots  $ & $\cdots$ & $\cdots $\\
\cline{1-8}\hline
\end{tabular}
\end{center}

In summary, we have suggested a method for implementing 
quantum logic operations between the BQ and CQ beyond the LD limit. 
In the controlled operations, the BQ acts as the control qubit.
It is worthwhile to point out that the BQ is really not an additional
qubit in the quantum computer because one can not perform any
single-qubit operation on the BQ. However, the BQ can serve as an
intermediary to perform the quantum logic operations between different CQs.
This will be the subject of the next section.
 
\section{Universal gates for quantum computation with two-level 
trapped cold ions beyond LD limit}

The ion trap quantum computer consists of a string of ions stored in a
linear radio-frequency trap and cooled sufficiently. The motion of the ions,
which are coupled together due to the Coulomb force between them, is quantum
mechanical in nature. Each qubit is formed by two
internal levels of an ion. The ions are
sufficiently separated to be addressed by different laser beams\cite{Ste},
i.e., each ion can be illuminated individually.
The communication and logic operations between qubits are performed
by using laser pulses sequentially to excite or de-excite quanta of the
collective vibration (i.e., the shared phonon) modes, which act as the BQ.
Following Monroe $et$ $al$\cite{Mor2},
only the lowest vibration mode of the ion string, i.e., the harmonic motion
of the center-of-mass, is considered. In the interaction picture defined
by the unitary operator
$$
\hat{U}_{0}^{N}(t)=\exp [-i\omega t(\hat{a}^{\dagger }\hat{a}
+1/2)]\prod_{j=1}^{N}\exp (-it\delta _{j}\hat{\sigma}_{j,z}/2),
$$
the Hamiltonian of the system takes the form
\begin{eqnarray}
\hat{H}=\frac{\hbar }{2}\sum_{j=1}^{N}\Omega _{j}\{\hat{\sigma}_{j,+}\exp
[i\eta _{j}(\hat{a}+\hat{a}^{\dagger })]+H.c.\},
\end{eqnarray}
with $\Omega _{j}\equiv \Omega ,\ \eta _{j}\equiv \eta $. 
We have shown in the above section that simple rotations on
a single CQ and controlled operations between a CQ and the BQ
can be realized exactly by applying suitable laser pulses.
In the following we show how to implement universal quantum gates
involving many CQs assisted by their common BQ.

\subsection{Operation of multi-CQ}

Operations on different CQs can be performed by applying different laser
beams to different ions. As a consequence,
rotation operations on 
different CQs can be performed individually and synchronously. 
For example, for a quantum register with  $N$ trapped cold ions,
it is easily proven that the uniform superposition state of $N$-CQ register
$$
|\Psi_0\rangle=\frac{1}{\sqrt{2}}\sum _{i=0}^{N-1}c_i|i\rangle, \hspace{0.2cm}c_i=\pm 1,
$$
which is the computational initial state for almost all quantum algorithms, can be easily prepared by applying $N$ resonant pulses with equal initial phases $\pi/2$ (or $3\pi/2$) and durations
$\Omega_{m,m}t=\pi/4$ to $N$ trapped ions synchronously.
We note that the CQs in such an operator are not correlated with
each other except by sharing a common phonon mode. For realizing
quantum computation with the $N$-ion quantum register,
we shall construct the universal two-qubit controlled gates between
different CQs.

\subsection{$C^Z$ logic gate between different trapped ions}

The $C^Z$ logic operation between $i$th ion and $j$th ion
\begin{eqnarray}
\hat{C}^Z_{c_ic_j}=|g_{i}\rangle |g_{j}\rangle \langle g_{i}|\langle
g_{j}|+|g_{i}\rangle |e_{j}\rangle \langle g_{i}|\langle
e_{j}|+|e_{i}\rangle |g_{j}\rangle \langle e_{i}|\langle
g_{j}|-|e_{i}\rangle |e_{j}\rangle \langle e_{i}|\langle e_{j}|,
\end{eqnarray}
means that if the first CQ (control qubit) is in the state $|g_i\rangle $,
the operation has no effect, whereas if the control qubit is in the state $%
|e_i\rangle $, the state of the second CQ (target qubit) is rotated by the
Pauli operator $\hat{\sigma}_z$. To realize this operation we consider
the quantum evolution of a three-qubit system (i.e., two CQs and the
BQ initially in the state $|0\rangle$) driven by an off-resonant laser pulse. 
Using three-step red-sideband
pulses with the same frequency
$\omega_L=\omega_0-\omega$ to the $i$th, $j$th and again the
$i$th ions sequentially, we have the following dynamical evolutions:
\begin{eqnarray}
\left\{
\begin{array}{lll}
|0\rangle |g_{i}\rangle |g_{j}\rangle &\rightarrow &|0\rangle |g_{i}\rangle
|g_{j}\rangle , \\
\\
|0\rangle |g_{i}\rangle |e_{j}\rangle &\rightarrow &B_{1}|0\rangle
|g_{i}\rangle |e_{j}\rangle +B_{2}|0\rangle |e_{i}\rangle |g_{j}\rangle
+B_{3}|1\rangle |g_{i}\rangle |g_{j}\rangle ,\\
\\
|0\rangle |e_{i}\rangle |g_{j}\rangle &\rightarrow &C_{1}|0\rangle
|g_{i}\rangle |e_{j}\rangle +C_{2}|0\rangle |e_{i}\rangle |g_{j}\rangle
+C_{3}|1\rangle |g_{i}\rangle |g_{j}\rangle ,\\
\\
|0\rangle |e_{i}\rangle |e_{j}\rangle &\rightarrow &D_{1}|1\rangle
|g_{i}\rangle |e_{j}\rangle +D_{2}|1\rangle |e_{i}\rangle |g_{j}\rangle
+D_{3}|2\rangle |g_{i}\rangle |g_{j}\rangle +D_{4}|0\rangle |e_{i}\rangle
|e_{j}\rangle,
\end{array}
\right.
\end{eqnarray}
with
$$
%\begin{eqnarray}
\left\{
\begin{array}{l}
B_{1} =\cos \Omega _{0,1}t'_{2},\hspace{0.2cm} B_{2}=-e^{i(\phi' _{2}-\phi' _{3})}\sin
\Omega _{0,1}t'_{2}\sin \Omega _{0,1}t'_{3},\hspace{0.2cm} B_{3}=-e^{i\phi' _{2}}\sin \Omega
_{0,1}t'_{2}\cos \Omega _{0,1}t'_{3};  \\
\\
C_{1} =-e^{i(\phi' _{1}-\phi' _{2})}\sin \Omega _{0,1}t'_{1}\sin \Omega
_{0,1}t'_{2},\hspace{0.2cm}C_{2} =\cos \Omega _{0,1}t'_{1}\cos \Omega
_{0,1}t'_{3}-e^{i(\phi'_1-\phi'_3)}\sin\Omega _{0,1}t'_1\cos\Omega _{0,1}t'_2\sin\Omega _{0,1}t'_3,\\
\\
C_{3} =-e^{i\phi'_{1}}\sin \Omega _{0,1}t'_{1}\cos \Omega _{0,1}t'_{2}\cos
\Omega _{0,1}t'_{3}-e^{i\phi' _{3}}\cos \Omega _{0,1}t'_{1}\sin \Omega _{0,1}t'_{3};\\
\\
D_{1} =-(e^{i\phi'_{1}}\sin \Omega _{0,1}t'_{1}\cos \Omega _{1,2}t'_{2}\cos
\Omega _{0,1}t'_{3}+e^{i\phi'_{3}}\cos \Omega _{0,1}t'_{1}\cos \Omega
_{0,1}t'_{2}\sin \Omega _{0,1}t'_{3}), \\
\\
D_{2} =-e^{i\phi' _{2}}(\cos \Omega _{0,1}t'_{1}\sin \Omega _{0,1}t'_{2}\cos
\Omega _{1,2}t'_{3}+e^{i(\phi' _{1}-\phi' _{3})}\sin \Omega _{0,1}t'_{1}\sin \Omega
_{1,2}t'_{2}\sin \Omega _{1,2}t'_{3}), \\
\\
D_{3} =e^{i\phi' _{2}}(e^{i\phi' _{1}}\sin \Omega _{0,1}t'_{1}\sin \Omega
_{1,2}t'_{2}\cos \Omega _{1,2}t'_{3}+e^{i\phi'_{3}}\cos \Omega _{0,1}t'_{1}\sin
\Omega _{0,1}t'_{2}\sin \Omega _{1,2}t'_{3}),\\
\\  
D_{4} =\cos \Omega _{0,1}t'_{1}\cos \Omega _{0,1}t'_{2}\cos \Omega _{0,0}t'_{3}
-e^{i(\phi'_{1}-\phi'_{3})}\sin \Omega _{0,1}t'_{1}\cos \Omega
_{1,2}t'_{2}\sin \Omega _{0,1}t'_{3}. 
\end{array}
\right.
%\end{eqnarray}
$$
Here $\phi'_{i}$ and $t'_{i}$ ($i=1,2,3$) are the phase and duration of the first, second, and third 
off-resonant pulses.
They should be set up properly for realizing the operation
$\hat{C}^Z_{c_ic_j}$. Obviously, if the LD parameter $\eta$ and the duration of 
the second off-resonant pulse are set up according to Eq.\,(9)
to implement the logic operation $\hat{C}^Z_{c_jb}$ between the BQ and the
$j$th CQ, the coefficients in Eq.\,(20) reduce to
$$
%\begin{eqnarray}
\left\{
\begin{array}{l}
B_{1} =1, \hspace{0.2cm} B_{2}=B_{3}=0;\\
\\
C_{1} =0,\hspace{0.2cm}C_{2} =\cos \Omega _{0,1}t'_{1}\cos \Omega
_{0,1}t'_{3}-e^{i(\phi'_1-\phi'_3)}\sin\Omega_{0,1}t'_1\sin\Omega_{0,1}t'_3,\\
\\
C_{3} =-e^{i\phi'_{1}}\sin \Omega _{0,1}t'_{1}\cos
\Omega _{0,1}t'_{3}-e^{i\phi'_{3}}\cos \Omega _{0,1}t'_{1}\sin \Omega _{0,1}t'_{3};\\
\\
D_{1} =e^{i\phi'_{1}}\sin \Omega _{0,1}t'_{1}\cos
\Omega _{0,1}t'_{3}-e^{i\phi' _{3}}\cos \Omega _{0,1}t'_{1}\sin \Omega _{0,1}t'_{3}), \\
\\
D_{2} =D_{3} =0,\hspace{0.2cm}  
D_{4} =\cos \Omega _{0,1}t'_{1}\cos \Omega
_{0,1}t'_{3}+e^{i(\phi' _{1}-\phi' _{3})}\sin \Omega _{0,1}t'_{1}\sin \Omega _{0,1}t'_{3}. 
\end{array}
\right.
%\end{eqnarray}
$$
These coefficients further become
$$
B_2=B_3=0,\hspace{0.2cm} B_1=1,\hspace{0.4cm}
C_1=C_3=0,\hspace{0.2cm} C_2=1,\hspace{0.4cm}
D_1=D_2=D_3=0,\hspace{0.2cm} D_4=-1,
$$
if $\phi'_1,\phi'_3$ and $t'_1,t'_3$ are further set up to satisfy the conditions
\begin{eqnarray}
\left\{
\begin{array}{l}
\phi' _{1} = \phi' _{3}\pm 2k\pi, k=0,1,2,...,\\
\\
\cos \Omega _{0,1}(t'_{1}+t'_{3})=1,\hspace{0.4cm} \cos \Omega
_{0,1}(t'_{1}-t'_{3})=-1.
\end{array}
\right.
\end{eqnarray}
This means that, once the LD parameter is set up properly, a $\hat{C}^Z_{c_jb}$ 
gate operation, surrounded by two off-resonant pulses addressing the $i$th CQ, yields the exact
$C^Z$ logic operation between the $i$th and $j$th ions, i.e.
\begin{equation}
\hat{R}_{c_ib}(\phi'_3,t'_3)\hat{C}^Z_{c_jb}\hat{R}_{c_ib}(\phi'_1,t'_1)\longrightarrow 
\hat{C}^Z_{c_ic_j}\otimes |0\rangle\langle 0|.
\end{equation}
The BQ remains in
its initial state after the operation. Eq.\,(21) implies that
\begin{eqnarray}
\Omega _{0,1}t'_{1}=(k+k^{\prime })\pi -\frac{\pi }{2},\hspace{0.3cm} 
\Omega _{0,1}t'_{3}=(k-k^{\prime })\pi +\frac{\pi }{2};\hspace{0.3cm}k, k'=1,2,3....
\end{eqnarray}
This shows that, once the LD parameter $\eta $ is set up properly,
the $C^Z$ logic operation (19) between the $i$th and the $j$th CQs can be
realized exactly by sequentially applying three-step red-sideband pulses
with adjustable durations.
The procedure for realizing this gate can be summarized as follows:\vspace{0.2cm}

a). A red-sideband pulse with $\Omega _{0,1}t'_{1}=3\pi/2$ is applied to
the $i$th ion;

b). A red-sideband pulse with $\Omega _{0,1}t'_{2}=2\pi$ is applied
to the $j$th ion;

c). A red-sideband pulse with $\Omega _{0,1}t'_{3}=\pi/2$,
whose phase equals to that of the first one, is applied to the $i$th ion
again.\vspace{0.2cm}

Comparing to Cirac-Zoller's proposal\cite{CZ1}, we note that the same number pulses for 
implementing the $C^Z$ logic operation between different ions are required.
An obvious advantage of the present method is that no auxiliary atomic level
is needed. Thus laser pulses with different polarizations are not
required.

\subsection{$C^N$ logic operation between different trapped ions}

As another universal two-qubit gate for ion trap quantum computer, the $C^N$
logic operation between a pair of CQs (e.g., the $i$th and $j$th ions),
\begin{eqnarray}
\hat{C}^N_{c_ic_j}=|g_{i}\rangle |g_{j}\rangle \langle g_{i}|\langle
g_{j}|+|g_{i}\rangle |e_{j}\rangle \langle g_{i}|\langle
e_{j}|+|e_{i}\rangle |g_{j}\rangle \langle e_{i}|\langle
e_{j}|+|e_{i}\rangle |e_{j}\rangle \langle e_{i}|\langle g_{j}|,
\end{eqnarray}
means that if the first CQ (control qubit) is in the state $|g_i\rangle $,
the operation has no effect, whereas if the control qubit is in the state
$|e_i\rangle $, the state of the second CQ (target qubit) undergoes a NOT
operation.
In Ref.\,\cite{Mor2} Monroe $et$ $al$ showed that if the $C^N$ gate (15)
between the target qubit $j$ and the BQ is surrounded by two extra
operations, which map and reset the state of the control qubit $i$ onto the
state of the BQ, the $C^N$ gate operation (24) may be carried out.
We now give an alternative method to realize the $C^N$ logic operation (24)
by making use of the two-qubit operation $\hat{C}^Z_{c_ic_j}$
and the single-qubit operation $\hat{r}_c(m,\phi,t)$.
When the BQ is in the state $|0\rangle $, we know from section II that
the rotation on the $j$th CQ
\begin{eqnarray}
\hat{r}_{c_j}(0,\phi, t)=\cos \Omega _{0,0}t|g_{j}\rangle \langle
g_{j}|-ie^{i\phi }\sin \Omega _{0,0}t|g_{j}\rangle \langle e_{j}|-ie^{-i\phi
}\sin \Omega _{0,0}t|e_{j}\rangle \langle g_{j}|+\cos \Omega _{0,0}t|e_{j}\rangle
\langle e_{j}|,
\end{eqnarray}
can be carried out by applying a single resonant laser pulse to the $j$th
ion. The state of the BQ is not changed during this operation. If the $\hat{C}^Z_{c_ic_j}$ gate is surrounded by the operations $\hat{r}_{c_j}(0,\phi''_1, t''_{1})$
and $\hat{r}_{c_j}(0,\phi''_3,t''_{3})$, we have
\begin{eqnarray}
\hat{r}_{c_j}(0,\phi''_3,t''_{3})\hat{C}^Z_{c_ic_j}\hat{r}_{c_j}(0,\phi''_1,t''_{1}) &=&B_{0000}|g_{i}\rangle |g_{j}\rangle \langle g_{i}|\langle
g_{j}|+B_{0001}|g_{i}\rangle |g_{j}\rangle \langle g_{i}|\langle
e_{j}|+B_{0010}|g_{i}\rangle |e_{j}\rangle \langle g_{i}|\langle g_{j}|+
\nonumber \\
&&B_{0011}|g_{i}\rangle |e_{j}\rangle \langle g_{i}|\langle
e_{j}|+B_{1100}|e_{i}\rangle |g_{j}\rangle \langle e_{i}|\langle
g_{j}|+B_{1101}|e_{i}\rangle |g_{j}\rangle \langle e_{i}|\langle e_{j}|+ \\ \nonumber
&&B_{1110}|e_{i}\rangle |e_{j}\rangle \langle e_{i}|\langle
g_{j}|+B_{1111}|e_{i}\rangle |e_{j}\rangle \langle e_{i}|\langle e_{j}|,
\end{eqnarray}
with
$$
%\begin{eqnarray}
\left\{
\begin{array}{l}
B_{0000} =\cos \Omega _{0,0}t''_{3}\cos \Omega _{0,0}t''_{1}-e^{i(\phi'' _{3}-\phi''
_{1})}\sin \Omega _{0,0}t''_{3}\sin \Omega _{0,0}t''_{1},\\
\\
B_{0001}=-i(e^{i\phi''
_{1}}\cos \Omega _{0,0}t''_{3}\sin \Omega _{0,0}t''_{1}+e^{i\phi'' _{3}}\sin \Omega _{0,0}
t''_{3}\cos \Omega _{0,0}t''_{1}), \\
\\
B_{0010} =-i(e^{-i\phi'' _{3}}\sin \Omega _{0,0}t''_{3}\cos \Omega
_{0,0}t''_{1}+e^{-i\phi'' _{1}}\cos \Omega _{0,0}t''_{3}\sin \Omega _{0,0}t''_{1}),\\
\\
B_{0011}=\cos \Omega _{0,0}t''_{3}\cos \Omega _{0,0}t''_{1}-e^{-i(\phi'' _{3}-
\phi''_{1})}\sin \Omega _{0,0}t''_{3}\sin \Omega _{0,0}t''_{1}, \\
\\
B_{1100} =\cos \Omega _{0,0}t''_{3}\cos \Omega _{0,0}t''_{1}+e^{i(\phi'' _{3}-\phi''_{1})}
\sin \Omega _{0,0}t''_{3}\sin \Omega _{0,0}t''_{1},\\
\\
B_{1101}=-i(e^{i\phi''_{1}}\cos \Omega _{0,0}t''_{3}\sin \Omega _{0,0}t''_{1}-e^{i\phi'' _{3}}\sin \Omega
_{0,0}t''_{3}\cos \Omega _{0,0}t''_{1}), \\
\\
B_{1110} =-i(e^{-i\phi'' _{3}}\sin \Omega _{0,0}t''_{3}\cos \Omega _{0,0}t''_{1}-e^{-i\phi'' _{1}}\cos \Omega _{0,0}t''_{3}\sin \Omega _{0,0}t''_{1}),\\
\\
B_{1111}=-\cos \Omega _{0,0}t''_{3}\cos \Omega _{0,0}t''_{1}-e^{-i(\phi'' _{3}-\phi''
_{1})}\sin \Omega _{0,0}t''_{3}\sin \Omega _{0,0}t''_{1}.
\end{array}
\right.
%\end{eqnarray}
$$
Here, $\phi''_1, t''_1$ and $\phi''_3, t''_3$ are the phases and durations of the
resonant pulses applied to perform two rotations on the target
CQ, i.e. the $j$th ion. Similarly, one can easily prove that,
if $\phi''_1,\phi''_3$ and $t''_1,t''_3$ satisfy
\begin{eqnarray}
\left\{
\begin{array}{l}
\phi''_3=\phi''_1\pm 2k\pi, k=0,1,2,...,\\
\\
\Omega _{0,0}t''_1=(p+p'-3/4)\pi,\hspace{0.3cm}\Omega _{0,0}t''_3=(p-p'+3/4)\pi, \hspace{4mm}{\mbox for }
\hspace{2mm}\phi_1=\pi/2\pm 2k\pi, p,p'=1,2,3,...;\\
\\
\Omega _{0,0}t''_1=(p+p'-1/4)\pi,\hspace{0.3cm}\Omega _{0,0}t''_3=(p-p'+1/4)\pi, \hspace{4mm}{\mbox for }
\hspace{2mm}\phi_1=3\pi/2\pm 2k\pi,
\end{array}
\right. 
\end{eqnarray}
the coefficients in Eq. (26) become
$$
B_{0000}=B_{0011}=B_{1101}=B_{1110}=1,\hspace{0.3cm}
B_{0001}=B_{0010}=B_{1100}=B_{1111}=0.
$$
This means that the exact $\hat{C}^Z_{c_ic_j}$
surrounded by two resonant 
pulses applied to the $j$th CQ can give rise to the exact $C^N$ logic
operation between the $i$th and $j$th CQs, i.e.,
\begin{equation}
\hat{r}_{c_j}(0,\phi''_3,t''_{3})\hat{C}^Z_{c_ic_j}\hat{r}_{c_j}(0,\phi''_1,t''_{1})\longrightarrow \hat{C}^N_{c_ic_j}.
\end{equation}
The durations of two applied resonant pulses are determined again by
Eq.\,(18).
We summarize the process for realizing the logic operation
$\hat{C}^N_{c_ic_j}$ as follows:\vspace{0.2cm}

a). A resonant pulse with initial phase $\pi/2$ ($3\pi/2$) and
$\Omega _{0,0}t''_1=5\pi/4$($7\pi/4$) is applied to the target CQ (the
$j$th ion);\vspace{0.2cm}

b). Three-step red-sideband pulses are sequentially used to implement $C^Z$
gate (19) between the $i$th and $j$th CQs;\vspace{0.2cm}

c). Another resonant pulse with initial phase $\pi/2$ ($3\pi/2$)
and $\Omega _{0,0}t''_3=3\pi/4$ ($\pi/4$) is applied to the target $j$th CQ.\vspace{0.2cm}

In summary, we have given a method to implement
arbitary one-qubit rotations on any single CQ
and the $C^N$ or $C^Z$ gate between
any pair of CQs with trapped cold ions.
These operations form a universal set. Any unitary operations on
an arbitrary number of CQs can be expressed as a the 
composition of the elements in the set. We note that the initial state
of the BQ in the above discussion is always assumed to be in its ground state
$|0\rangle$. One can prove that, if the BQ is initially in the excited state,
e.g. $|1\rangle$, $|2\rangle$, $etc.$, universal two-qubit gates
between different ions can not be realized by a one-quantum excitation process.
Implementing quantum computation by making use of the
multiquantum (multiphonon) interaction between the CQ and BQ 
will be discussed in a future publication.  
 
\section{Conclusions}

We have demonstrated the possibility of performing ion trap quantum
computation with two-level ions beyond the LD limit.
The method for realizing the exact universal quantum gates, i.e., single-qubit
rotations and two-qubit controlled logic operations, has been discussed
in detail. The present scheme is not limited to small values of the
LD parameter and does not require an extra atomic level.
Therefore, laser pulses with different polarizations are not needed.
There are two requirements for our scheme. First, the LD parameter $\eta$ must be set up accurately for realizing
the exact $C^Z$ logic operation between the internal and external
degrees of freedom. Second, the durations of all applied laser pulses must be adjusted accurately.
We have shown
that the $C^Z$ logic operation between the external and internal states 
can be realized by using a single off-resonant pulse.
This $C^Z$ gate $\hat{C}^Z_{cb}$ surrounded by two resonant pulses
with suitable durations gives rise to the exact corresponding $C^N$ logic gate.
We have further shown that the $C^Z$ logic operation between different ions
in a trap can be implemented by using a three-step sequence of off-resonant
pulses. The second off-resonant pulse is used to realize the
$C^Z$ logic operation between the internal and external states
of the target ion, and the first and third ones are applied to the control ion. 
This $C^Z$ gate $\hat{C}^Z_{c_i c_j}$ surrounded
by two resonant pulses with suitable durations applied
to the target ion, yields the corresponding $C^N$ logic operation
between two ions.

Compared to other methods for realizing the exact $C^N$ logic operation
on the ion-trap quantum register,
the present scheme has some advantages.  Although the same
number of pulses are needed for realizing the $C^N$ logic gate
as that reported in Refs.\,\cite{CZ1},\cite{CC} and \cite{Mor1},
the auxiliary atomic level and LD approximation are not required
in the present scheme. Our approach to realizing the $C^N$ gate
is also different from that in Ref.\,\cite{Mor2} and \cite{LG},
where a controlled operation between the BQ and CQ, which is
equivalent to the exact $C^N$ gate only under local transformation,
was realized by applying a single resonant pulse.
In the present scheme the exact $C^N$ logic gate is implemented
by an off-resonant pulse surrounded by two resonant ones.
We hope that the present scheme will be useful for future
experiments. 

\vspace{0.2cm}

\section*{acknowledgments}

This work was supported by the Natural Science Foundation of China, the
Special Funds for Major State Basic Research Project (grant No. 200683),
the Shanghai Municipal Commission of Science and Technology, and the
Shanghai Foundation for Research and Development of Applied Materials. One 
of us (Wei) is grateful to Prof. J. Q. Liang for helpful discussions.

\end{document}